\documentclass[
    ,final            
  ]
  {aipproc}

\layoutstyle{6x9}

\begin{document}

\title{GAMMA RAY BURSTS AND RADIO LOUD ACTIVE GALACTIC NUCLEI}

\author{Gabriele Ghisellini}{
  address={Osservatorio Astronomico di Brera, via Bianchi 46 I--23807 Merate Italy}
}

\begin{abstract}
We believe that the radiation we receive from Gamma-Ray Bursts
(GRBs) and radio loud Active Galacti Nuclei (AGNs) originates from
the transformation of bulk relativistic motion into random energy.
Mechanisms to produce, collimate and accelerates the jets
in these sources are uncertain, and it may be fruitful to
compare the characteristics of both class of sources in search
of enlightening similarities.
I will present some general characteristics of radio loud AGNs and 
GRBs such as their bulk Lorentz factors and the power of their jets.
I will also discuss the way in which the energy in bulk
relativistic motion can be transformed into beamed radiation, and
consider the possibility that both classes of sources
can work in the same way, namely by an intermittent release of
relativistic plasma at the base of the jet: shells ejected with
slightly different velocities collide at some distance from the
central engine, dissipating part of their kinetic energy, and
keeping the rest to power the extended radio lobes (in AGNs) or
to produce the afterglow (in GRBs).

\end{abstract}

\maketitle


\section{INTRODUCTION}

Radio loud Active Galactic Nuclei (AGNs) and Gamma Ray Bursts (GRBs)
have very little in common, at first sight.
GRB are flashes of $\gamma$--ray radiation, likely flagging the
birth of a stellar size black hole, while radio--loud AGNs, even if 
remarkable for their rapid variability, live for hundreds millions
years, producing spectacular and Mpc--size jets and radio lobes,
and are powered by supermassive black holes.
On the other hand, in both classes the emitting plasma is moving
at relativistic bulk speeds, and the radiation we see is likely
the result of the transformation of part of this well ordered
kinetic energy into random energy and then into radiation.
Furthermore, there are strong evidences that also GBRs have 
collimated jets.
And finally, consider that the dynamical timescale for a GRB
should be of the order the light travel time to cross the gravitational
radius, i.e. $R_{\rm g}/c\sim 10^{-4}M_1$ seconds, where $M_1$ is the mass
of the black hole in units of tens of solar masses.
A burst with a duration of 10 seconds therefore lasts for $10^5$ 
dynamical times: it can be a quasi steady process (for a $10^9$ solar mass
black hole, this time is equivalenth to 30 years).
What we naively consider an ``explosion" is instead a long event.

In both classes of sources we have non--thermal particles and
magnetic field, suggesting that non--thermal radiation processes
are the main contributors to the radiation we see.
This radiation, being produced by plasma in relativistic motion,
is strongly beamed in the velocity direction, and we have evidences
that also in GRBs the emitting fireball is collimated in a cone,
i.e. a ``jet".
For these reasons it is instructive to compare them looking for
similarities and differences, to see if their physics is similar.
In the following I will briefly discuss some of the basic facts
of blazars and GRBs, and discuss the possibility that, at the origin
of their phenomenology, there is a common engine.

\section{Blazars}

\begin{figure}
\includegraphics[width=1.1\textwidth,height=0.48\textheight]{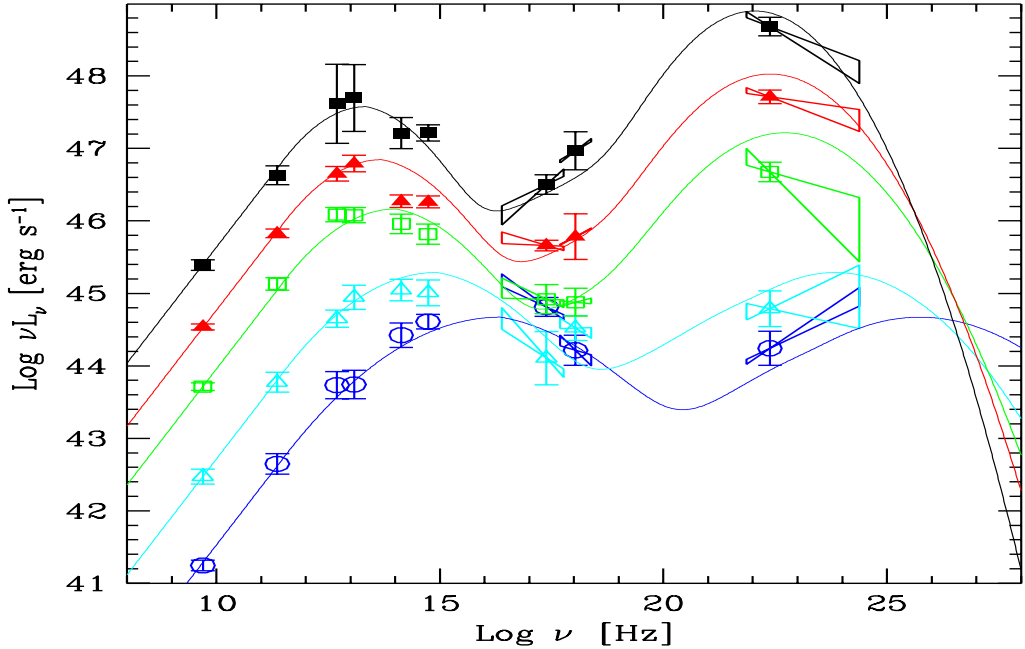}
\caption{The blazar sequence: note the shift of the peak energies as
the total power changes. From Fossati et al. (1998) and Donato et al. (2002).}
\label{sequence}
\end{figure}

\noindent{\bf Bulk Lorentz factor ---}
The best evidences for bulk relativistic motion
in blazars come from VLBI observations of
knots of radio emission moving superluminally.
Apparent speeds up to $\beta_{\rm app} =30 h^{-1}$  
are measured (see e.g. Jorstad et al. 2001),
at least in those blazars that are powerful $\gamma$--ray emitters, 
indicating even larger Lorentz factors $\Gamma$ (note that $\Gamma$ 
must always be larger than $\beta_{\rm app}$).
Indications of a large degree of beaming come also from
spectral fitting, especially of low powerful, TeV emitting, blazars
where beaming factors $\delta>20$ are derived (Tavecchio et al. 1998). 
There are also indications that the jet could be structured, 
with a fast ``spine" surrounded by a slower ``layer"
(Laing 1993; Chiaberge et al. 2001; Giroletti et al. 2003), explaining for example
why the non--thermal radiation
of the core of radio--galaxies is not as faint as predicted if
the plasma is moving at the same large speed and if they are 
observed at large angles with respect to the jet axis.


\vskip 0.2 true cm
\noindent
{\bf The blazar sequence ---}
Fossati et al. (1998), collecting data from three complete sample
of blazars, demonstrated that the SED is controlled by the bolometric
observed luminosity, with both peaks shifting at smaller frequences when 
increasing the luminosity (see Fig. \ref{sequence}).
Furthermore, the dominance of the high energy peak increases when
increasing the bolometric luminosity
(but this latter inference was based on the few 
low power BL Lacs detected by EGRET).
This ``blazar sequence" can be explained by a different degree of
radiative cooling: in powerful blazars electrons cool faster,
producing a break in the electron distribution functions
at smaller and smaller energies when increasing the total
(radiation plus magnetic) energy density in the comoving frame
(Ghisellini et al. 1998). 

\vskip 0.2 true cm
\noindent
{\bf Jet power ---}
The radio lobes of radio--galaxies and
blazars are a sort of calorimeter: the power required to feed them
can be calculated dividing the total energy of a radio lobe
by its lifetime (estimated from spectral aging or from advance motion). 
This estimate has been done, among others, by Rawling \& Saunders (1991): 
they find an average power ranging from $10^{43}$--$10^{44}$ erg s$^{-1}$ 
for FR I radiogalaxies to $10^{46}$--$10^{47}$ erg s$^{-1}$ for FR II
radiogalaxies and radio--loud quasars.

One can also calculate the power carried by the jet by
inferring its density through modeling the observed SED
and requiring that the jet carries at least the particles
and the magnetic field necessary to make the radiation we see.
This has been done on the pc scale by Celotti \& Fabian (1993),
on sub--pc scale (the $\gamma$--ray emitting zone) 
by Celotti \& Ghisellini (2003, see also Ghisellini 2003), 
and on the hundreds of kpc scale (the X--ray jets
seen by Chandra) by Celotti, Ghisellini \& Chiaberge (2001)
and Tavecchio et al. (2000).
These studies suggest large values of the power transported
by the jet and require the presence of a dynamically dominating 
proton component (see also arguments by Sikora \& Madejski 2000).



\section{GAMMA RAY BURSTS}

The main breaktrough in GRB science was the precise
localization of some of them made possible by the coded 
mask of the wide field camera of {\it Beppo}SAX, which
in turn made possible the prompt follow up in
X--rays, optical and radio.
Then it was possible to measure the redshift and end
a decade long discussion about the galactic or cosmological
origin of GRBs.
Up to now, about 30 redshifts of GRBs have been measured.
Apart from the controversial case of GRB 980425, possibly
associated with the nearby SN 1998bw (at $z\sim 0.008$),
all other redshifts are within the 0.17--4.5 range.
A particularly useful updated link with all the relevant information
about bursts with good localization is maintained by Jochen Greiner at:
{\tt www.aip.de/{$\sim$}jcg/grbgen.html}~~.

\vskip 0.2 true cm
\noindent
{\bf Duration ---} 
The majority of GRBs lasts for more than 2 seconds, while
about one third is shorter.
All information derived from the precise localization of GRBs refer to 
{\it long} bursts.
The bimodality of the distribution of their duration 
is confirmed by the associated spectral shape,
since short bursts, on average, appear harder than long GRBs. 
The light curve of GRBs is erratic and sometimes highly
variable: spikes as short as a fraction of a millisecond have 
been detected (see Shaefer \& Walker 1999).
The extremely short timescales we observe demand 
large Lorentz factors, and the fact that the spikes
at early and late times of the prompt emission have
similar timescales (i.e. their duration does not increase) are
major proofs against external shocks (see below) causing
the prompt emission of GRBs (Fenimore, Ramirez--Ruiz \& Wu 1999).

\vskip 0.2 true cm
\noindent
{\bf Spectra of the prompt emission ---} 
The spectra of GRBs are very hard, with a 
peak (in a $E$--$E F_E$ plot) at 
an energy $E_{peak}$ of a few hundreds keV.
Some bursts have been detected at very large $\gamma$--rays energies 
($>$ 100 MeV) by the EGRET instrument (see the review by Fishman 
\& Meegan 1995, and references therein).

\vskip 0.2 true cm
\noindent
{\bf The GRB--Supernova connection ---}
GRB 030329 is certainly associated with the
supernova 2003dh (see e.g. Stanek et al. 2003).
This burst, exceptionally bright because close by ($z\sim$0.17),
will probably be a Rosetta stone for GRB science.
We now have quite a secure confidence that the progenitors
of GRBs are massive stars, most likely exloding as SN Ic.
What remains to do is to find if there is a lag between the
SN and the GRB explosion in some bursts, as envisaged by the
SupraNova scenario (Vietri \& Stella 1998).

\vskip 0.2 true cm
\noindent
{\bf Iron lines ---}
For a few bursts, there are evidence for large amounts of X--ray 
line emitting material around the site of the explosion.
The detection of emission features (albeit with relatively low
significance) in the afterglow spectra of 
GRBs some hours after the GRB event poses strong constraints
on the properties of the line--emitting material
(see Lazzati 2003 for a recent review).

\vskip 0.2 true cm
\noindent
{\bf Jet breaks ---}
Assume that the burst is collimated within a cone of semiaperture
$\theta$.
Assume also that, initially, the bulk Lorentz factor of the fireball is
such that $1/\Gamma <\theta$.
Because of relativistic aberration, 
the observer (which is within the cone defined by $\theta$) 
will receive light only from a cone of aperture angle $1/\Gamma$.
This leads to the estimate of how the received flux varies in time.
If the fireball is spherical, this will continue as long as the
motion is relativistic.
But if the fireball is collimated, there is a time when $1/\Gamma$
becomes comparable to $\theta$.
After this time the observed solid angle will remain constant,
and then there will be a change in the slope of the light curve.
An {\it achromatic} break is predicted (``jet break"),
This break allows to estimate $\theta$ and then to correct the
isotropic values of the energetics of GRBs.
Frail et al. (2001) in this way obtain the remarkable result
that despite the ``isotropic" energy values differ by some orders
of magnitude, the corrected values are all very similar 
and cluster around a value of a few times $10^{51}$ erg.

\vskip 0.2 true cm
\noindent
{\bf Polarization ---}
GRB 021206 was serendipitously observed by RHESSI (a satellite for 
solar studies), showing a prompt emission (in the hard X--rays)
linearly polarized at the extraordinary level of $(80\pm20)$\%
(Coburn \& Boggs 2003). 
Polarimetric observations were performed for several afterglows
in the optical, finding always a moderate (but non--zero) linear
polarization at the 1--3\% level (see Covino et al. 2003).

\vskip 0.2 true cm
\noindent 
{\bf Bulk Lorenz factors ---}
If the source is moving relativistically, then the observed photon energies
are blueshifted, and the typical angles (as observed in the lab frame)
between photons are smaller, decreasing the probability for them to interact.
Bulk Lorentz factors $\Gamma>100$ are required to avoid strong 
suppression of high energy $\gamma$--rays due to photon--photon collisions.

There is a second argument demanding for strong relativistic motion,
concerning the very fast observed variability.
In fact the size associated with one millisecond is $R\sim 3\times 10^7$cm,
which is much too small to be optically thin.
To match the observed timescales with the size at which 
the fireball becomes transparent ($R_t\sim 10^{13}$ cm)
we need a Doppler contraction of time given approximately by 
$ct_{var} \sim R_t (1-\beta)$, yielding $\Gamma \sim 400$.

\begin{figure}
\includegraphics[height=0.35\textheight]{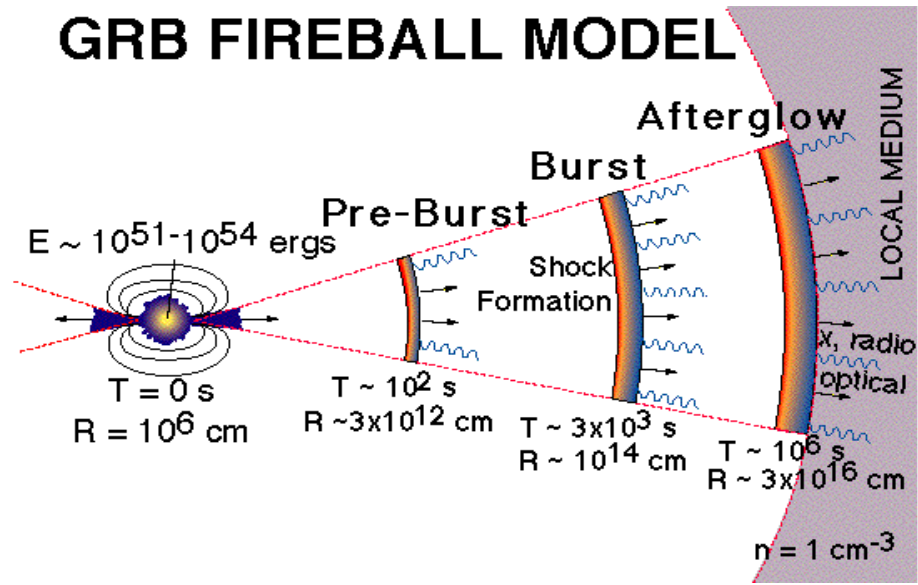}
\caption{The ``standard" fireball model for GRBs.}
\end{figure}

\subsection{The fireball}
\noindent
If there is a huge release of energy in a small volume,
no matter in which form the energy is initially injected,
a quasi--thermal equilibrium (at relativistic temperatures)
between matter and radiation is reached,
with the formation of electron--positron pairs accelerated to bulk
relativistic speeds by the high internal pressure.
This is a {\it fireball} (Cavallo \& Rees 1978).
See Fig. 2.
When the temperature of the radiation (as measured in the comoving
frame) drops below $\sim$50 keV the pairs annihilate 
faster than the rate at which they are produced.
But the presence of even a small amount of barions, corresponding to 
only $\sim 10^{-6}~ M_\odot$,
makes the fireball opaque to Thomson scattering: the internal radiation
thus continues to accelerate the fireball until most of its initial 
energy has been converted into bulk motion. 
After this phase the fireball expands at a constant speed and at some point 
becomes transparent.

If the central engine does not produce a single pulse, but works 
intermittently, it can produce many shells (i.e. many fireballs)
with slightly different Lorentz factors.
Late but faster shells can catch up early slower ones,
producing shocks which give rise to the observed burst emission.
In the meantime, all shells interact with the interstellar medium, and 
at some point the amount of swept up matter is large enough to decelerate
the fireball and produce other radiation which can be identified with 
the afterglow emission observed at all frequencies.

This is currently the most accepted picture for the burst 
and afterglow emission,
and it is called the internal/external shock scenario 
(Rees \& M\'esz\'aros 1992; 
Rees \& M\'esz\'aros 1994; 
Sari \& Piran 1997).
According to this scenario, the burst emission is due to
collisions of pairs of relativistic shells (internal shocks), while
the afterglow is generated by the collisionless shocks produced by shells
interacting with the interstellar medium (external shocks).

\section{Internal shocks for blazars}

Rees (1978) was the first to propose the internal shocks idea to transport
energy from the nucleuos to the outer jet of M87 (a famous AGN).
As mentioned above, this idea then became ``standard" for GRBs.
But it can work equally well, if not better, for blazars.
For powerful blazars, in fact, we require that most of the power 
is not dissipated and transformed into radiation, but kept to feed
the huge radio lobes.
Internal shocks are not very efficient in transforming bulk into random
energies (the shells are both relativistic: it is only the {\it relative}
kinetic energy that can be dissipated).
This is a problem for GRBs, due to the theoretical desire to limit the
total energetic amount of GRBs, but it is welcome in the blazar field.
Therefore Ghisellini (1999) and Spada et al. (2001) applied these ideas
to blazars, finding that the SED and the general behavior of blazars could
be well reproduced by this scenario.
In addition, one can find a simple explanation of why the jet is
dissipationless in its first $\sim 100$ Schwarzchild radii:
this is the minimum distance required by a shell to catch up
the previous one.

With respect to GRBs, there is an important difference: in the case
of blazars the shell cannot be accelerated, initially, by its
internal radiation pressure: this would cause some visible effects
on their SED, which are not observed: the shells must be initially cold.

\section{Matter or magnetically dominated jets?}

The internal shock scenario requires the jets to be matter dominated.
The role of the Poynting flux, at the scales of 
the dissipation regions, should be minor.
On the other hand Blandford (2003) and Lyutikov \& Blandford (2003)
propose an alternative scenario both for GRBs and radio loud AGNs,
in which the jet is magnetically dominated at all scales.
The radiation could be produced, in this case, by reconnection.
Since the magnetic field is the dominant jet energy carrier,
in this scenario the matter is less important, and the jet could
even be made by electron--positron pairs required to produce
the radiation we see, with a negligible proton component.

We have seen above that the bolometric luminosity of {\bf blazars} 
is often dominated by the high energy peak: if this is due
to the inverse Compton process, then the relative importance
of the synchrotron component must be minor.
Therefore this ``Compton dominance" implies a modest
role of the magnetic field in the emitting region, 
to limit the synchrotron emitted power.
This conclusion relies on the assumption that the high energy
peak of blazars is due to inverse Compton.
It could be, instead, again due to the synchrotron process, by
ultrarelativistic electrons and positrons, resulting from 
electromagnetic cascades involving relativistic protons.
But in this case the cascading process would result also in an
overproduction of X--rays, at the ``valley" between the two peaks
(see Ghisellini 2003).

For {\bf GRBs} the issue is much more controversial (and interesting)
because of two recent observational results:
1) in some burst the spectrum of the prompt emission can be 
fitted by a blackbody, at least for the first seconds
(Ghirlanda et al. 2003);
2) the prompt emission of GRB 021206 was strongly linearly polarized
(Coburn \& Boggs 2003).
Some blackbody radiation is expected in the standard, hot fireball,
scenario: the radiation responsible for the acceleration of
the fireball can escape when the fireball becomes
transparent, and it can well be blackbody in shape, in 
close analogy with the fossil cosmic background radiation.
But the very large polarization, instead, would point 
towards a very important, dominant, and very well ordered
magnetic field.

\section{The same engine?}
The primary energy source of GBRs and blazars could well be the same.
The main store of energy is the spin of the black hole:
for a maximally rotating Kerr hole, one can extract the 29\% of its
total mass--energy, amounting to $\sim 5\times 10^{53}M/M_\odot$ erg.
The problem is how to extract this energy sufficiently fast
(especially in the case of GRBs).
One promising way to extract this energy is the Blandford \& Znajek (1977) 
process, in which the rotational energy of a Kerr black hole 
can be extracted by a magnetic field surrounding the hole providing
a source of power:
\begin{equation}
L_{\rm BZ}\,  \sim \, 10^{51}\, \left({a\over m}\right)^2\,
\left({M_{\rm BH} \over 10\, M_\odot}\right)^2
\left({B\over 10^{14}{\rm G}}\right)^2\,\, {\rm erg\, s^{-1}} \,  
\sim \, \left({ a\over m}\right)^2  (3R_s)^2 U_B c
\end{equation}
where $(a/m)$ is the specific black hole angular momentum 
($\sim 1$ for maximally rotating black holes), $R_s$ is the Schwarzchild 
radius and $U_B=B^2/(8\pi)$.
The duration of a typical GRB could be associated to the duration 
of the accretion process (a few seconds for long bursts).
To sustain such large magnetic fields, the torus surrounding them
should be very dense, but this is only natural, given the fact that it
is the core of the progenitor star not yet collapsed into the hole.

Even if this mechanism is purely magnetic, suggesting that the
shells are born ``cold" also in GRBs, it is not clear if the shell
can remain cold: the shell could be initially purely magnetic and cold,
but soon a fraction of the transported energy could be converted into
hot pairs and trapped radiation, forming a classical hot fireball.
On the other hand, for blazars, we do have observational contraints
suggesting that this does not occur, but we also have other convincing
evidence that the shells in blazars have magnetic fields
below equipartition. The situation is rather puzzling, and this is
an open issue.

One should also consider that short GRBs could be powered by the same 
basic mechanism even if their progenitors are two merging neutron stars
(with the possible help of $\sim 10^{51}$ ergs in neutrinos).

\section{CONCLUSIONS}

GRBs last for tens of thousands of dynamical times, and are not single explosions,
as supernovae, even if the association between GRBs and supernovae is now certain.
It is very likely that they are collimated, and their radiation is certainly beamed.
Their power can exceed $10^{50}$ erg s$^{-1}$ in $\gamma$--rays
even accounting for collimation, and the total emitted energy is of the order 
of $10^{52}$ ergs.
Being so luminosus, albeit for a short time, they are the best torchlights we have
to illuminate the far universe.
Since they are associated with massive stars, there is the hope to study,
through them, Pop III stars and the re--ionization phases of the universe, at
redshifts as large as 15--20.
The basic physics of GRBs may be similar to the physics of relativistic jets
in general, and therefore share many aspects with blazars, even if they are
obviously more extreme.
For both classes of sources we may see, at action, the more efficient engine
that nature invented to produce radiation, more efficient than accretion.
That this is the case is already clear considering those blazars that although 
having powerful jets, do not show any sign of thermal emission coming from 
accretion, such as lineless BL Lacs.
But  that jets can be orders of magnitude more powerful than accretion becomes
dramatically evident with GRBs.





\end{document}